\documentclass[12pt]{article}

\usepackage{amsmath, amssymb, amsfonts}
\usepackage{graphicx}
\usepackage{booktabs}
\usepackage{natbib}
\usepackage{url}
\usepackage{setspace}
\usepackage{geometry}
\usepackage{hyperref}
\usepackage{lineno}

\usepackage{authblk}

\hypersetup{
  colorlinks=true,
  linkcolor=blue,
  citecolor=blue,
  urlcolor=blue
}

\title{\textbf{Temporal dependence in exposure and hazard-based infectious disease interventions}}

\author{
Hiroyasu Ando$^{1}$\thanks{Corresponding author: hiro1999@g.ucla.edu}, A. James O’Malley$^{2,3}$, Akihiro Nishi$^{4}$}

\date{}

\begin{document}

\maketitle

\footnotetext[1]{Department of Statistics \& Data Science, University of California, Los Angeles, CA.}
\footnotetext[2]{Department of Biomedical Data Science, Geisel School of Medicine at Dartmouth, NH.}
\footnotetext[3]{The Dartmouth Institute for Health Policy and Clinical Practice, Geisel School of Medicine at Dartmouth, NH.}
\footnotetext[4]{Department of Epidemiology, University of California, Los Angeles, CA.}

\begin{abstract}
In randomized controlled trials (RCTs) of infectious disease interventions, it is well recognized that unmeasured individual heterogeneity at baseline can induce selection bias over time, thereby complicating the interpretation of the estimated hazard ratio. The present study examines a simplified setting: RCTs consisting of homogeneous participants, with no individual heterogeneity at baseline. However, even in such an apparently ideal setting, selection bias can emerge over time due to temporal dependence in exposure, a realistic feature of infectious disease transmission. In this study, we mathematically characterize the mechanism underlying this bias and quantitatively evaluate its magnitude. Our results show that this bias should be recognized as an issue in both the design and interpretation of RCTs of infectious disease interventions.
\end{abstract}

\begin{center}
\end{center}

\section{Introduction}

In randomized controlled trials (RCTs) of infectious disease interventions, it is well recognized that unmeasured individual heterogeneity at baseline can induce selection bias over time, thereby complicating the interpretation of the estimated hazard ratio ($\widehat{\text{HR}}$). This selection bias gives rise to multiple concerns: (i) it calls into question whether $1-\widehat{\text{HR}}$ provides an unbiased estimate of the average per-exposure effect \citep{O'Hagan2014bias}, and (ii) it can lead to violations of the proportional hazards assumption (PHA) \citep{stensrud2025hazard}. A key contributing factor is a fundamental feature of such trials: individual-level exposure data are typically unobserved \citep{Stensrud2023ve, O'Hagan2014bias}.

Following \cite{O'Hagan2014bias}, in the absence of individual-level exposure data, the conditions required to avoid this selection bias can be summarized as follows:
(A) the average per-exposure effect is constant over time;
(B) the average per-exposure effect does not depend on past exposure history;
(C) there is no unmeasured heterogeneity in susceptibility to infection; and
(D) there is no confounding affecting the effect of exposure.

Motivated by these considerations, the present study examines a simplified setting: an RCT consisting of homogeneous participants, with no individual heterogeneity at baseline and without individual-level exposure data. Under this setting, the four conditions described above \citep{O'Hagan2014bias} are satisfied. Moreover, because the per-exposure effect coincides with the average per-exposure effect, issues related to the non-collapsibility of the hazard ratio \citep{dumas2025hazard} can be set aside.

However, even in such an apparently ideal setting, exposure heterogeneity can emerge over time due to temporal dependence in exposure, a realistic feature of infectious disease transmission. Consequently, selection bias may arise despite the absence of baseline heterogeneity, and the interpretability of $\widehat{\text{HR}}$ may again become problematic. In this study, we mathematically characterize the mechanism underlying this bias and quantitatively evaluate its magnitude.

\section{Mechanism}\label{sec2}

\begin{figure} 
\centerline{
\includegraphics[width=\textwidth]{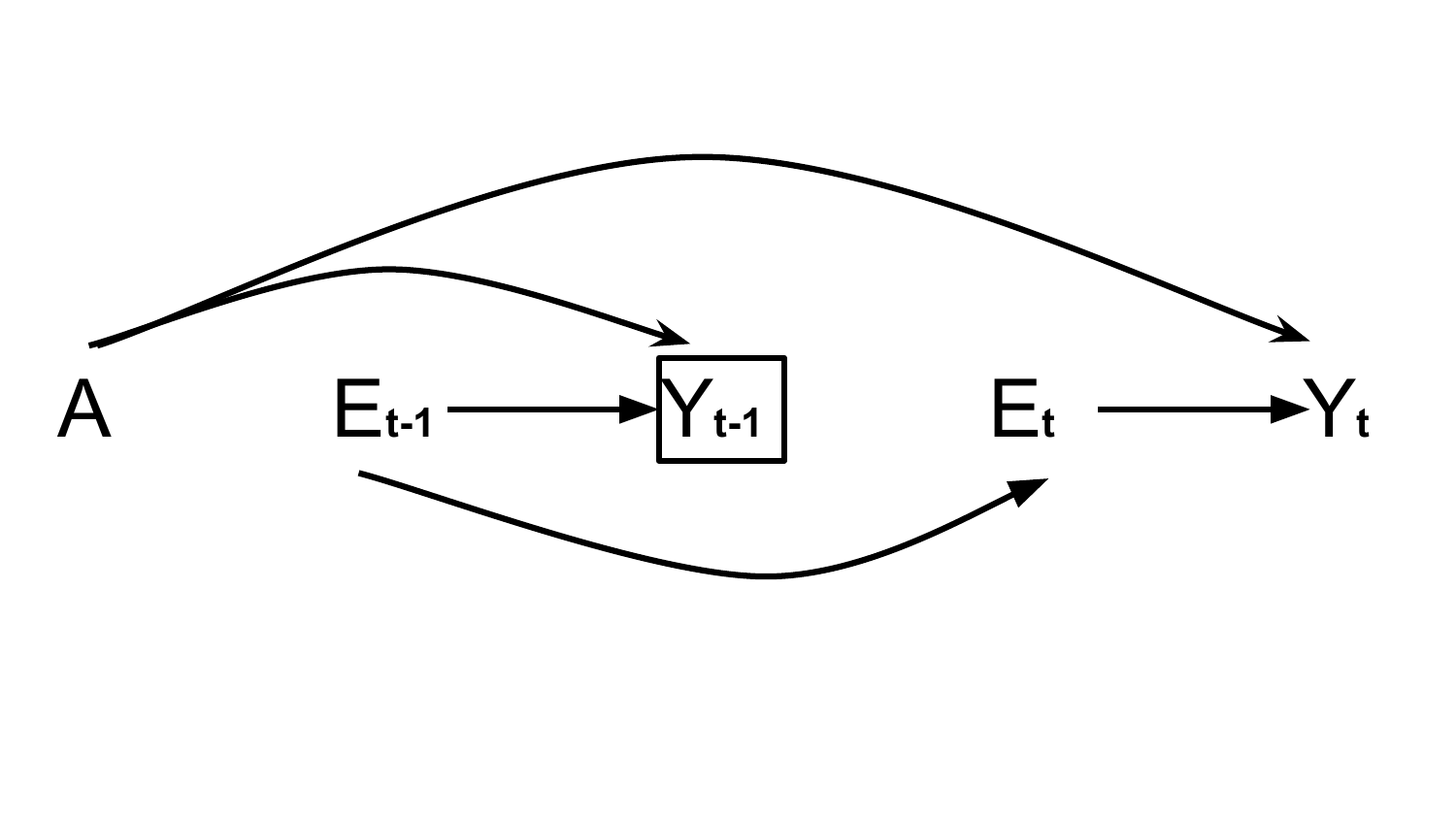}}
\caption{Causal diagram for a double-blind randomized trial with $\mathbf{A} \rightarrow \boxed{\mathbf{Y_{t-1}}} \leftarrow \mathbf{E_{t-1}} \rightarrow \mathbf{E_{t}} \rightarrow \mathbf{Y{_t}}$}
\label{fig3}
\end{figure}

In this section, we describe a mechanism through which selection bias emerges over time due to temporal dependence in exposure, even in seemingly ideal RCTs of infectious disease interventions consisting of homogeneous participants and lacking individual-level exposure data.

First, we consider typical RCTs in which exposure is not controlled by the investigator, in contrast to challenge studies \citep{killingley2011challenge, roestenberg2009challenge}, where exposure is deliberately controlled. In such settings, individual-level exposure data are not observed. 

Second, we assume that participants exhibit no individual heterogeneity; that is, all participants share the same exposure probability, per-exposure infection probability, and per-exposure effect of the intervention. We consider a leaky vaccine effect \citep{smith1984ve}, in which the per-exposure effect is defined as the proportional reduction in the per-exposure infection probability and is constant over time. Under this setting, the per-exposure effect coincides with the average per-exposure effect.

Although vaccine effects can be modeled using either a leaky or an all-or-nothing model \citep{smith1984ve, Kahn2018bias}, the latter is incompatible with the homogeneous setting considered here. We therefore adopt the leaky vaccine model.

In Fig.~\ref{fig3}, let $\mathbf{A} \in \{0,1\}$ denote treatment assignment at time $t=0$. Let $\mathbf{Y_t} \in \{0,1\}$ denote infection status at time $t$, with $\mathbf{Y_0}=0$, and let $\mathbf{E_t} \in \mathbb{Z}_{\ge 0}$ denote the number of exposure events at time $t$.

To estimate the (average) per-exposure effect at time $t$, we consider the discrete-time HR:
\[
\frac{\mathbb{P}(\mathbf{Y_t}=1 \mid \mathbf{Y_{t-1}}=0,\mathbf{A}=1)}
{\mathbb{P}(\mathbf{Y_t}=1 \mid \mathbf{Y_{t-1}}=0,\mathbf{A}=0)},
\]
where $\mathbb{P}(\mathbf{Y_t} = 1 \mid \mathbf{Y_{t-1}} = 0, \mathbf{A} = j)$ is the probability of infection at time $t$ under treatment ($j=1$) or no treatment ($j=0$), given no prior infection at time $t-1$.
Here, conditioning on $\mathbf{Y_{t-1}}=0$ is represented by $\boxed{\mathbf{Y_{t-1}}}$ in Fig.~\ref{fig3}.

Under the setting described so far, the conditions outlined in \cite{O'Hagan2014bias} are satisfied, and $1-\widehat{\text{HR}}$ would be expected to provide an unbiased estimate of the (average) per-exposure effect. The four conditions are as follows: (A) the average per-exposure effect is constant over time; (B) the average per-exposure effect does not depend on past exposure history;
(C) there is no unmeasured heterogeneity in susceptibility to infection; and (D) there is no confounding affecting the effect of exposure.

However, we now focus on temporal dependence in exposure, represented by the direct path $\mathbf{E_{t-1}} \rightarrow \mathbf{E_t}$ in Fig.~\ref{fig3}. Then, conditioning on $\mathbf{Y_{t-1}}=0$ opens the following path: 
\[
\mathbf{A} \rightarrow \boxed{\mathbf{Y_{t-1}}}
\leftarrow \mathbf{E_{t-1}} \rightarrow \mathbf{E_t} \rightarrow \mathbf{Y_t}.
\]
Consequently, the quantity $1-\text{HR}$ deviates structurally from the (average) per-exposure effect.

The path $\mathbf{E_{t-1}} \rightarrow \mathbf{E_t}$ has rarely been made explicit in existing DAGs for RCTs of infectious disease interventions \citep{O'Hagan2014bias, Stensrud2023ve}. However, temporal dependence in exposure naturally arises in real-world transmission dynamics.

For example, if a participant is exposed at time $t-1$ (i.e., $\mathbf{E_{t-1}} > 0$), the infectious source is likely to remain infectious at time $t$, thereby increasing the probability of subsequent exposure (i.e., $\mathbf{E_t} > 0$). For instance, during the early phase of the COVID-19 pandemic, the average infectious period was estimated to be approximately three days \citep{He2020covid}. In settings where participants repeatedly interact in environments such as households, workplaces, or schools, exposure events tend to occur over time rather than independently.

\section{Modeling}\label{sec3}

\subsection{Infectious window assumption}\label{subsec3.1}

In this section, we formulate a mathematical model for RCTs of infectious disease interventions based on the mechanism described in Section~\ref{sec2}. We assume that participants are drawn from a homogeneous population with no individual heterogeneity at baseline. That is, all participants share the same exposure probability, per-exposure infection probability, and per-exposure effect; consequently, the per-exposure effect coincides with the average per-exposure effect. In addition, individual-level exposure data are assumed to be unobserved.

\paragraph{Study design}
At time $t=0$, each participant is randomly assigned to either the treatment group ($\mathbf{A}=1$) or the placebo group ($\mathbf{A}=0$), and is subsequently followed over discrete time points $t=0,1,2,\dots$.

\paragraph{Exposure process}

At each time point, an individual may initiate an event referred to as an infectious window. An infectious window represents a latent event corresponding to sustained contact with a single infectious source over multiple time points. Once initiated, it generates one unit of exposure at each consecutive time point, including the time of initiation. Let the duration of an infectious window be represented by a positive integer-valued random variable $I \in \mathbb{N}_{>0}$.

We assume that the initiation of an infectious window occurs independently for each individual at each time point with probability $m$, where $m \approx 0$, and is identical across individuals and time. At each time point, an individual can initiate at most one new infectious window; however, infectious windows initiated at different time points may overlap. As a result, multiple units of exposure may occur simultaneously at a given time point.

For example, if an infectious window initiated at time $t-1$ persists over multiple time points and a new infectious window is initiated at time $t$, then the total number of exposures at time $t$ is two.

\paragraph{Infection event} 

Let $\mathbf{E}_t \in \mathbb{Z}_{\ge 0}$ denote the total number of exposure units at time $t$. Conditional on $\mathbf{E}_t = E$, infection at time $t$ is assumed to occur with probability:
\[
1 - (1-p)^E,
\]
where $p \in (0,1)$ is the per-exposure infection probability, assumed to be identical across individuals and time. This corresponds to assuming that each exposure at a given time point results in an independent Bernoulli trial with success probability $p$. This type of epidemic model is well established and widely used \citep{Nishi2020covid, herrera2016disease}.

Once individuals become infected, they are permanently removed from the risk set, and reinfection is not allowed. 

\paragraph{Intervention}

We assume a leaky vaccine \citep{smith1984ve, Kahn2018bias}, in which the per-exposure effect is defined as the proportional reduction in the per-exposure infection probability ($p$). Specifically, in the treatment group ($\mathbf{A}=1$), the per-exposure infection probability ($p$) is reduced to $p (1-v)$, where $v \in (0,1)$ represents the (average) per-exposure effect and is assumed to be identical across individuals and over time. Therefore, given $E$ exposures at time $t$, the probability of infection is
\[
1 - \{1 - p (1-v)\}^E.
\]

\subsection{$1-\text{HR}$ and the (average) per-exposure effect}\label{subsec3.2}

In this section, we consider RCTs of infectious disease interventions consisting of homogeneous participants without individual-level exposure data (see Section~\ref{subsec3.1}). The goal is to formalize the relationship between $1-\text{HR}$ and the (average) per-exposure effect ($v$).

Let $\mathbf{A} \in \{0,1\}$ denote treatment assignment at time $t=0$. Let $\mathbf{Y_t} \in \{0,1\}$ denote infection status at time $t$, with $\mathbf{Y_0}=0$. Let $\mathbf{E_t} \in \mathbb{Z}_{\ge 0}$ denote the number of exposure events at time $t$. The discrete-time HR at time $t$ is defined as:
\begin{equation}
\text{HR} = \frac{\mathbb{P}_\theta(\mathbf{Y}_{t}=1
\mid
\mathbf{Y}_{t-1}=0,\mathbf{A}=1)}{\mathbb{P}_\theta(\mathbf{Y}_{t}=1
\mid
\mathbf{Y}_{t-1}=0,\mathbf{A}=0)}. \label{hr}
\end{equation}

The target estimand is the (average) per-exposure effect, $v \in (0,1)$, defined as:
\begin{equation}
v
=
1-
\frac{
\mathbb{P}_\theta(\mathbf{Y}_{t}=1
\mid
\mathbf{Y}_{t-1}=0,\mathbf{A}=1,\mathbf{E}_t=1)
}{
\mathbb{P}_\theta(\mathbf{Y}_{t}=1
\mid
\mathbf{Y}_{t-1}=0,\mathbf{A}=0,\mathbf{E}_t=1)
},\label{ve}
\end{equation}
Because infectious windows may persist over time, the number of exposure events at time $t$, $\mathbf{E}_t$ can take values as large as $t \in \mathbb{N}$. 

By the law of total probability, the hazard at time $t$ can be expressed as:
\begin{align}
&\mathbb{P}_\theta(\mathbf{Y}_{t}=1
\mid
\mathbf{Y}_{t-1}=0,\mathbf{A}) \nonumber \\
&= 
\sum_{e=1}^{t}
\mathbb{P}_\theta(\mathbf{Y}_{t}=1
\mid
\mathbf{Y}_{t-1}=0,\mathbf{A},\mathbf{E}_t=e)
\mathbb{P}_\theta(\mathbf{E}_t=e
\mid
\mathbf{Y}_{t-1}=0,\mathbf{A}).
\end{align}

Assuming that the initiation probability of infectious windows is sufficiently small ($m \approx 0$), the probability that multiple infectious windows overlap at the same time point is of higher order. Under this assumption, the hazard at time $t$ can be:
\begin{align}
&\mathbb{P}_\theta(\mathbf{Y}_{t}=1
\mid
\mathbf{Y}_{t-1}=0,\mathbf{A})
\nonumber\\
&=
\mathbb{P}_\theta(\mathbf{Y}_{t}=1
\mid
\mathbf{Y}_{t-1}=0,\mathbf{A},\mathbf{E}_t=1)
\mathbb{P}_\theta(\mathbf{E}_t=1
\mid
\mathbf{Y}_{t-1}=0,\mathbf{A})
+ O_A(m^2). \label{approx_h}
\end{align}

Combining \eqref{hr}, \eqref{ve} and \eqref{approx_h}:
\begin{align}
1-\text{HR}
&=
1-
\frac{
\mathbb{P}_\theta(\mathbf{Y}_{t}=1
\mid
\mathbf{Y}_{t-1}=0,\mathbf{A}=1)
}{
\mathbb{P}_\theta(\mathbf{Y}_{t}=1
\mid
\mathbf{Y}_{t-1}=0,\mathbf{A}=0)
}
\nonumber\\
&\to
1-
(1-v)
\frac{
\mathbb{P}_\theta(\mathbf{E}_t=1
\mid
\mathbf{Y}_{t-1}=0,\mathbf{A}=1)
}{
\mathbb{P}_\theta(\mathbf{E}_t=1
\mid
\mathbf{Y}_{t-1}=0,\mathbf{A}=0)
}
\quad (\text{as } m \to 0).\label{hr_ape}
\end{align}

Therefore, $1-\text{HR}$ systematically deviates from the  (average) per-exposure effect ($v$) through the factor:
\begin{equation}
\frac{
\mathbb{P}_\theta(\mathbf{E}_t=1
\mid
\mathbf{Y}_{t-1}=0,\mathbf{A}=1)
}{
\mathbb{P}_\theta(\mathbf{E}_t=1
\mid
\mathbf{Y}_{t-1}=0,\mathbf{A}=0)
}.
\label{exp_ratio}
\end{equation}

Thus, characterizing the behavior of this conditional exposure probability ratio in \eqref{exp_ratio} is essential for understanding the mechanism underlying the selection bias.

\subsection{Conditional exposure probability ratio}\label{subsec3.3}

In this section, we consider randomized controlled trials (RCTs) of infectious disease interventions consisting of homogeneous participants without individual-level exposure data (see Section~\ref{subsec3.1}). The goal of this section is to quantitatively characterize the conditional exposure probability ratio introduced in \eqref{exp_ratio}.

Let $\mathbf{A} \in \{0,1\}$ denote treatment assignment at time $t=0$. Let $\mathbf{Y_t} \in \{0,1\}$ denote infection status at time $t$, with $\mathbf{Y_0}=0$. Let $\mathbf{E_t} \in \mathbb{Z}_{\ge 0}$ denote the number of exposure events at time $t$. Let $\mathbf{X}=(X_1,\ldots,X_t)\in\{0,1\}^t$ denote a binary vector indicating whether an infectious window is initiated at each time point.

By the law of total probability, the conditional exposure probability at time $t$ can be written as:
\begin{align}
\mathbb{P}_\theta(\mathbf{E}_{t}=1 \mid \mathbf{Y}_{t-1}=0,\mathbf{A}=a)
=
\sum_{\mathbf{x}\in\{0,1\}^t}
\mathbb{P}_\theta(\mathbf{E}_{t}=1,\mathbf{X}=\mathbf{x} \mid \mathbf{Y}_{t-1}=0,\mathbf{A}=a).
\end{align}

Under the assumption that the initiation probability of infectious windows is sufficiently small ($m \approx 0$), the probability of multiple initiations ($\|\mathbf{X}\|_1 \ge 2$) is of higher order. Restricting attention to:
\[
\mathcal{W}
=
\left\{\mathbf{w}\in\{0,1\}^t:\|\mathbf{w}\|_1=1\right\},
\]
we obtain:
\begin{align}
&\mathbb{P}_\theta(\mathbf{E}_{t}=1 \mid \mathbf{Y}_{t-1}=0,\mathbf{A}=a)
\nonumber\\
&=
\sum_{\mathbf{w}\in\mathcal{W}}
\mathbb{P}_\theta(\mathbf{E}_{t}=1,\mathbf{X}=\mathbf{w} \mid \mathbf{Y}_{t-1}=0,\mathbf{A}=a)
+O^*_{A=a}(m^2).
\end{align}
For $\mathbf{A} = 0$, a direct calculation yields:
\begin{align}
\label{eq:Et1_placebo}
&\mathbb{P}_\theta(\mathbf{E}_{t}=1 \mid \mathbf{Y}_{t-1}=0,\mathbf{A}=0) \\
& =
\frac{m \cdot (1-m)^{t-1}}{1-O^{**}_{A=0}(m \cdot p)}
\sum_{s=0}^{t-1}(1-p)^s\,\mathbb{P}(\mathbf{I}>s)
+O^*_{A=0}(m^2),
\end{align}
where $\mathbb{P}(\mathbf{I}>s)$ denotes the probability that an infectious window initiated at time $t-s$ persists until time $t$. $p$ denotes the per-exposure infection probability. Similarly, for $\mathbf{A} = 1$:
\begin{align}
\label{eq:Et1_vax}
&\mathbb{P}_\theta(\mathbf{E}_{t}=1 \mid \mathbf{Y}_{t-1}=0,\mathbf{A}=1) \\
&=
\frac{m \cdot (1-m)^{t-1}}{1-O^{**}_{A=1}(m \cdot p \cdot (1-v))}
\sum_{s=0}^{t-1}\{1-p \cdot (1-v)\}^s\,\mathbb{P}(\mathbf{I}>s)
+O^*_{A=1}(m^2),
\end{align}
where $v$ denotes the (average) per-exposure effect.

From \eqref{eq:Et1_placebo}--\eqref{eq:Et1_vax}, as $m \to 0$, the conditional exposure probability ratio satisfies:
\begin{align}
\label{eq:ratio_limit}
\frac{\mathbb{P}_\theta(\mathbf{E}_{t}=1 \mid \mathbf{Y}_{t-1}=0,\mathbf{A}=1)}
{\mathbb{P}_\theta(\mathbf{E}_{t}=1 \mid \mathbf{Y}_{t-1}=0,\mathbf{A}=0)}
\to
\frac{
\sum_{s=0}^{t-1}\{1-p \cdot (1-v)\}^s\,\mathbb{P}(\mathbf{I}>s)
}{
\sum_{s=0}^{t-1}(1-p)^s\,\mathbb{P}(\mathbf{I}>s)
}. 
\end{align}

\paragraph{Conservative bound}

For an integer $R \in \{1, ..., t-1\}$, the right-hand side of \eqref{eq:ratio_limit} admits the following conservative approximation:
\begin{align}
\label{eq:R_lower_bound}
\frac{
\sum_{s=0}^{R-1}\{1-p \cdot (1-v)\}^s\, \mathbb{P}(\mathbf{I}>s)
}{
\sum_{s=0}^{R-1}(1-p)^s\, \mathbb{P}(\mathbf{I}>s)
} \;\le\;
\frac{
\sum_{s=0}^{t-1}\{1-p \cdot (1-v)\}^s\, \mathbb{P}(\mathbf{I}>s)
}{
\sum_{s=0}^{t-1}(1-p)^s\, \mathbb{P}(\mathbf{I}>s)
} \quad
\because \quad \frac{1-p \cdot (1-v)}{1-p} >  1. 
\end{align}
This implies that the deviation between $1-\text{HR}$ and the (average) per-exposure effect ($v$) can be conservatively evaluated (please see \eqref{hr_ape}).

\paragraph{Relaxing the infectious window assumption}

The bound in \eqref{eq:R_lower_bound} also admits an interpretation that relaxes the infectious window assumption. Specifically, let the initiation probabilities of infectious windows be $m_1,\dots,m_t$ over time, each being sufficiently small. If, in addition, the most recent $R \in \{1, ..., t-1\}$ time points satisfy:
\[
m_{t-R+1}=\cdots=m_{t}\approx 0,
\]
then the same conservative bound in \eqref{eq:R_lower_bound} can be obtained. Choosing a smaller $R$ yields a tighter lower bound (i.e., greater underestimation of the conditional exposure probability ratio), while allowing greater flexibility in accommodating time-varying initiation probabilities. Thus, the choice of $R$ reflects a trade-off between the strength of the modeling assumptions and the conservativeness of the bound.

\section{Results}\label{sec4}

\subsection{Simulations}\label{subsec4.1}

\begin{figure}
\centerline{
\includegraphics[width=\textwidth]{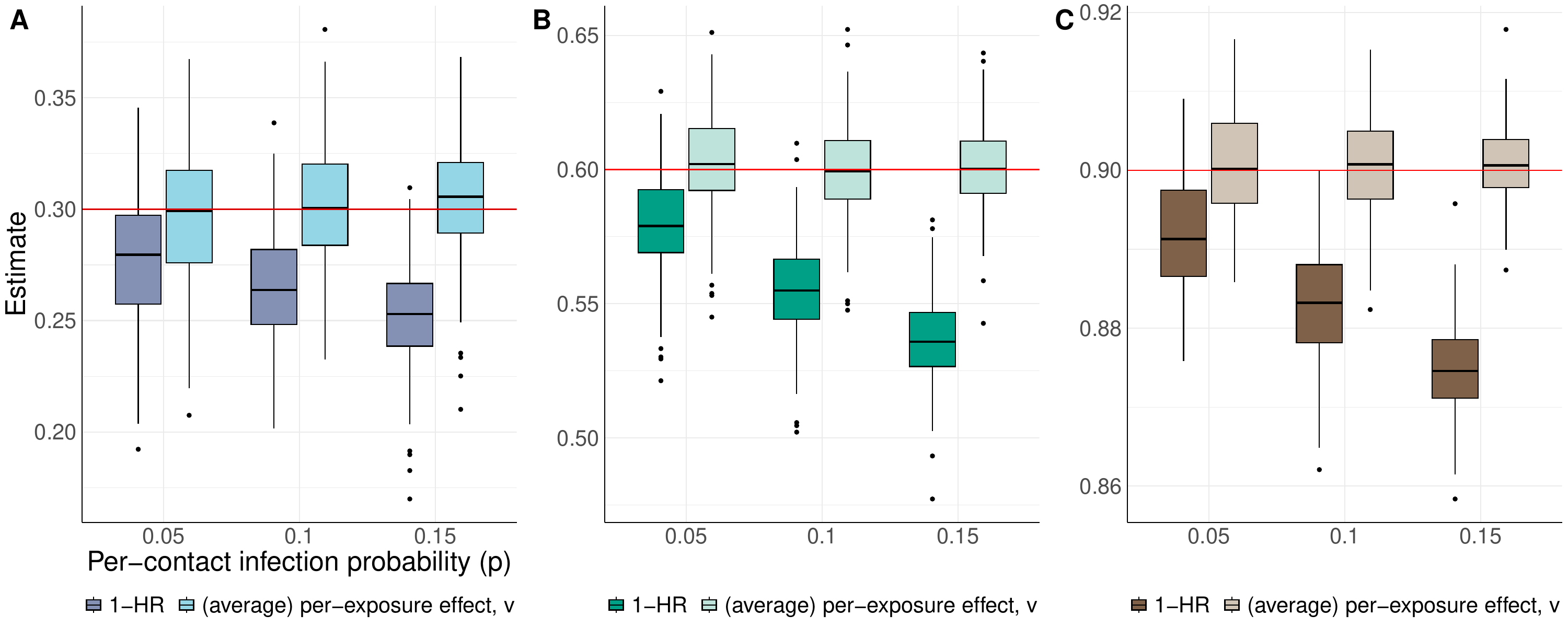}}
\caption{$1-\widehat{\text{HR}}$ and the estimated (average) per-exposure effect ($\hat{v}$) over the $200$ simulations}
\label{fig4}
\end{figure}

In this section, we consider RCTs of infectious disease interventions consisting of homogeneous participants without individual-level exposure data (see Section~\ref{subsec3.1}), and evaluate the relationship between $1-\text{HR}$ and the (average) per-exposure effect ($v$) derived in Sections~\ref{subsec3.2} and \ref{subsec3.3}. 

The simulation is designed in accordance with the setup in Section~\ref{subsec3.1}. The observation period consists of 180 time points (days), and $10{,}000$ individuals are randomly assigned to the treatment and placebo groups, with $5{,}000$ individuals in each group. The initiation probability of the infectious windows is set to $m = 0.05$, which is relatively large compared to the asymptotic limit $m \to 0$ and is intended as a sensitivity analysis. The duration of the infectious windows follows $\mathbf{I} \sim \textit{Geometric}(1/3)$, as in a previous COVID-19 study \citep{Nishi2020covid}. We consider nine simulation scenarios corresponding to all combinations of the per-exposure infection probability, $p \in \{0.05, 0.1, 0.15\}$ and the (average) per-exposure effect, $v \in \{0.3, 0.6, 0.9\}$.

We aim to conservatively correct the deviation between $1-\text{HR}$ and the (average) per-exposure effect ($v$) inherent from time $t=30$ onward. To this end, we first estimate HR using the Cox proportional hazards model \citep{Cox1972cox} based on simulated data from time $t=30$ days onward, and then obtain the conservative estimate of $v$, given pre-specified per-exposure infection probability ($p$) and the duration of an infectious window ($\mathbf{I}$) by solving:
\begin{align}
1-\text{HR}  
\frac{
\sum_{s=0}^{29}(1-p)^s\, \mathbb{P}(\mathbf{I}>s)}{
\sum_{s=0}^{29}\{1-p  (1-v)\}^s\, \mathbb{P}(\mathbf{I}>s)
} = v.
\label{hr_simu}
\end{align}
Equation~\eqref{hr_simu} extends the relationship in \eqref{hr_ape} together with the conservative bound in \eqref{eq:R_lower_bound}, corresponding to setting $R = 30$. 

This correction procedure is theoretically justified via the delta method. Specifically, under the transformation induced by Equation~\eqref{hr_simu}, we have:
\begin{align}
\sqrt{n}(\widehat{\text{HR}} - \text{HR}) \xrightarrow{d} \mathcal{N}(0, \sigma^2)
\quad \Longrightarrow \quad
\sqrt{n}(\widehat{v} - v) \xrightarrow{d} \mathcal{N}(0, \sigma^2_{v}).
\label{ve_adjusted}
\end{align}

Figure~\ref{fig4} presents the simulation results. In each panel, the red horizontal line represents the (average) per-exposure effect ($v$). Panel A (blue) corresponds to $v = 0.3$, Panel B (green) to $v = 0.6$, and Panel C (brown) to $v = 0.9$. These results show that $1-\widehat{\text{HR}}$ exhibits systematic bias with respect to $v$. In contrast, the adjusted estimator $\hat{v}$ obtained via Equation~\eqref{hr_simu} consistently reduces this bias across the range of the parameter settings. 

\subsection{Visualization}\label{subsec4.2}

\begin{figure} 
\centering
\includegraphics[width=1\textwidth]{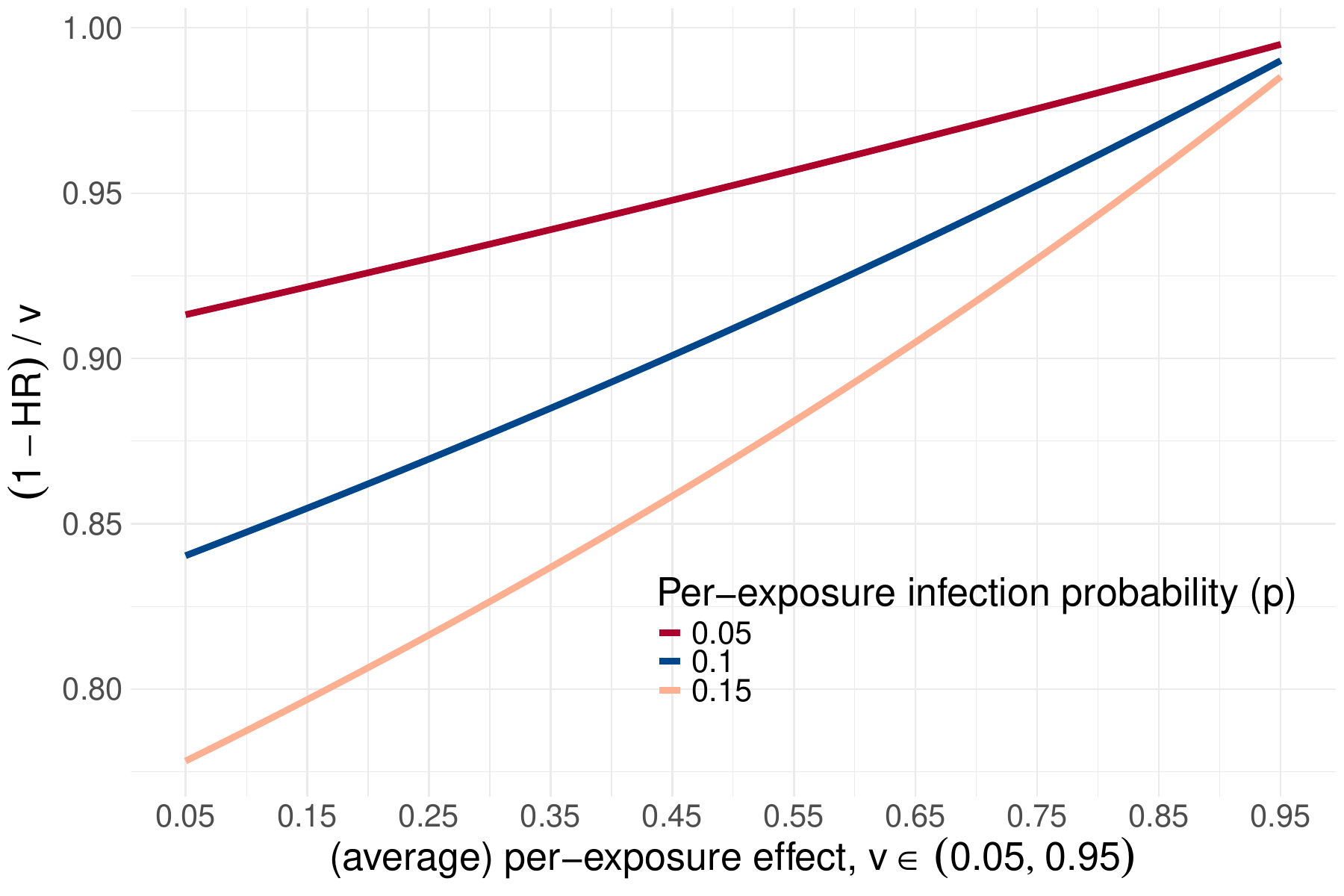}
\caption{$1-\text{HR}$ and the (average) per-exposure effect ($v$) at time $t=90$}
\label{fig5}
\end{figure} 

In this section, we consider RCTs of infectious disease interventions consisting of homogeneous participants without individual-level exposure data (see Section~\ref{subsec3.1}), and visualize the relationship between $1-\text{HR}$ and the (average) per-exposure effect ($v$) derived in Sections~\ref{subsec3.2} and \ref{subsec3.3}.

The relationship between $1-\text{HR}$ and the (average) per-exposure effect ($v$) at time $t$ is defined as:
\begin{align}
1-\text{HR} \cdot 
\frac{
\sum_{s=0}^{t-1}(1-p)^s\, \mathbb{P}(\mathbf{I}>s)}{
\sum_{s=0}^{t-1}\{1-p \cdot (1-v)\}^s\, \mathbb{P}(\mathbf{I}>s)
} = v.
\label{hr_graph}
\end{align}
This equation extends the relationship in \eqref{hr_ape} by incorporating the conditional exposure probability ratio in \eqref{eq:ratio_limit}.

In Figure~\ref{fig5}, x-axes is the (average) per-exposure effect, \( v \in (0.05, 0.95) \), and y-axes is the ratio $(1-\text{HR})/v$ based on Equation~\eqref{hr_graph}. As in the simulation, we assume the duration of the infectious windows, $\mathbf{I} \sim \text{Geometric}(1/3)$ and consider three values of the per-exposure infection probability, $p \in \{0.05, 0.1, 0.15\}$. The time point is fixed at $t=90$. For example, when \( v = 0.45 \) and \( p = 0.05 \), the ratio $(1-\text{HR})/v$ is approximately 0.95, indicating a $5\%$ deviation between $1-\text{HR}$ and $v$ at $t=90$. This deviation increases as $v$ decreases, and as $p$ increases.

\section{Discussion}\label{sec5}

We identified a mechanism by which temporal dependence in exposure induces selection bias over time in RCTs of infectious disease interventions consisting of homogeneous participants and lacking individual-level exposure data. This bias creates a  discrepancy between $1-\text{HR}$ and the (average) per-exposure effect, complicating the interpretation of $1-\text{HR}$ and potentially violating the proportional hazards assumption (PHA). Importantly, this phenomenon arises even in homogeneous populations, distinguishing it from conventional selection bias driven by unmeasured individual heterogeneity at baseline. Moreover, because exposure data are typically unavailable in practice, the bias identified in this study is likely to be present in many RCTs of infectious disease interventions.

As a consequence of this selection bias, $1-\widehat{\text{HR}}$ systematically underestimates the (average) per-exposure effect ($v$). For example, under the setting with per-exposure infection probability $p=0.15$, infectious window duration $\mathbf{I} \sim \text{Geometric}(1/3)$, and $v=0.5$, the discrepancy exceeds $10\%$ at time point $t=90$ (see Figure~\ref{fig5}). This magnitude of deviation is not negligible and may have a meaningful impact on the interpretation of intervention effects. For instance, during the COVID-19 pandemic, the FDA's minimum threshold for Emergency Use Authorization (EUA) was defined as a point estimate of $50\%$ effectiveness \citep{FDA2021EUA}, implying that even a small percentage difference in estimated effectiveness could affect policy decisions. Therefore, this bias should be recognized as an important issue in both the design and interpretation of RCTs of infectious disease interventions.

For the numerical evaluation of the bias, we assume that the initiation probability of infectious windows is very small ($m \approx 0$). However, we argue this assumption could be plausible in many settings. For example, in a COVID-19 study conducted in England, the prevalence between May 20 and June 7, 2021 was estimated to be at most approximately $0.0026$ \citep{riley2021react}. This implies that the initiation probability ($m$) at any given time point may be of a similar order. Furthermore, our simulation results remain robust even when a relatively large value of $m = 0.05$ is considered (see Section \ref{subsec4.1}).

Finally, on a set of simplifying assumptions (see Section~\ref{subsec3.1}), we emphasize the goal of this work is not to provide a practically implementable correction for $1-\text{HR}$, but rather to clarify the underlying mechanism of this bias, and to highlight its importance as a consideration in the design and interpretation of RCTs of infectious disease interventions.

\section*{Disclosure Statement}

AN is a consultant to Vacan, Inc. and obtained an honorarium from Taisho Pharmaceutical Co., Ltd., which had no role in the project.

\section*{Software and Data}
All analysis was done in the R environment (R version 4.3.2). The code and available data to reconstruct the analyses of this paper are available at \url{https://github.com/Ankoudon/ando2025survival}.

\section*{Study funding}
AN was supported by the National Institutes of Health ($\text{K01AI166347}$), the National Science Foundation ($\text{NSF}\#\text{2230125}$), and the Japan Science and Technology Agency ($\text{JPMJPR21R8}$). AJO was supported by the National Institutes of Health ($\text{P01AG019783, P20GM148278}$). The content is solely the responsibility of the authors and does not necessarily represent the official views of the funders.

\section*{Author Contributions}

HA discovered the bias. HA derived the formulas and methodologies, which were validated by AJO. HA, AN, and AJO wrote the manuscript. AN contributed laboratory
resources.

\bibliographystyle{chicago}

\bibliography{main}

\end{document}